\documentclass[pra,showpacs,superscriptaddress,longbibliography,twocolumn]{revtex4-1}

\usepackage{graphicx}
\usepackage{amsmath}

\usepackage{color}
\usepackage{cancel}
\usepackage{ulem}
\usepackage{float}

\begin{document}
\title{Cavity optomechanics with surface acoustic waves}

\author{Ayato~Okada}
\email{okada@qc.rcast.u-tokyo.ac.jp}
\author{Fumikazu~Oguro}
\affiliation{Research Center for Advanced Science and Technology (RCAST), The University of Tokyo, Meguro-ku, Tokyo 153-8904, Japan}
\author{Atsushi~Noguchi}
\affiliation{Research Center for Advanced Science and Technology (RCAST), The University of Tokyo, Meguro-ku, Tokyo 153-8904, Japan}
\affiliation{PRESTO, Japan Science and Technology Agency, Kawaguchi-shi, Saitama 332-0012, Japan}
\author{Yutaka~Tabuchi}
\affiliation{Research Center for Advanced Science and Technology (RCAST), The University of Tokyo, Meguro-ku, Tokyo 153-8904, Japan}
\author{Rekishu~Yamazaki}
\email{rekishu@qc.rcast.u-tokyo.ac.jp}
\affiliation{PRESTO, Japan Science and Technology Agency, Kawaguchi-shi, Saitama 332-0012, Japan}
\author{Koji~Usami}
\affiliation{Research Center for Advanced Science and Technology (RCAST), The University of Tokyo, Meguro-ku, Tokyo 153-8904, Japan}
\author{Yasunobu~Nakamura}
\affiliation{Research Center for Advanced Science and Technology (RCAST), The University of Tokyo, Meguro-ku, Tokyo 153-8904, Japan}
\affiliation{Center for Emergent Matter Science (CEMS), RIKEN, Wako, Saitama 351-0198, Japan}

\date{\today}

\begin{abstract}

We report a development of an electro-optomechanical system based on a surface acoustic wave (SAW), where a piezoelectric material with a large optoelastic susceptibility is used for the coupling of both a radio wave and optical light to the SAW.  In the optical domain, we exploit a tensorial nature of the optoelastic effect to show a polarization dependence of the photon-SAW interaction.  We discuss the construction of two-dimensional SAW focusing circuits for the coupling enhancement and the optical cavity enhanced photon-phonon scattering.  We estimate an optomechanical coupling rate $g_0$ of the system and discuss the future direction for the improvement of the coupling strength.

\end{abstract}


\maketitle

\section{Introduction}
Cavity optomechanical systems offer a novel way to optically manipulate the motion of mechanical oscillators ranging from nanoscale objects, such as nanobeam and microtoroidal resonators, to macroscopic objects such as a Fabry-P\'erot interferometer with suspended mirrors~\cite{Aspelmeyer2014}.  A prominent example is the laser cooling of the mechanical motion into the quantum ground state, achieved both in the microwave and optical regimes~\cite{Teufel2011,Chan2011}.  Maturity of these techniques and technologies may bring the quantum control of the macroscopic objects to a reality.  It is of fundamental interest to investigate quantum properties of macroscopic mechanical oscillators, such as entanglement and decoherence~\cite{Bose1997,Mancini1997,Bose1999}.  On the other hand, there are wide varieties of applications of the optomechanical systems.  The unprecedented precision in the displacement and force sensing using mechanical oscillators has been adopted for ultrasensitive measurements, such as the celebrated gravitational wave detector~\cite{PRL_LIGO}.  One of the unique features of the mechanical oscillators is its capability to couple to a wide range of electromagnetic waves, from radio waves to optical light.  The optomechanical systems are considered as one of the promising candidate for a microwave-to-light signal transducer~\cite{Bochmann2013,Andrews2014,Hisatomi2016}, which is an essential building block for a long-distant quantum network.

Many of the optomechanical systems utilize a `bulk' mode of a mechanical oscillator, including a drum mode of a thin membrane or a flexural mode of a wire.  On the other hand, there is a growing interest in using a surface mode, more specifically, a surface acoustic wave (SAW).  SAW is a propagating mechanical mode localized on the surface of the substrate.  The SAW can easily be excited and manipulated through patterned electrodes on a piezoelectric substrate or on a piezoelectric film.  A SAW resonator with patterned Bragg mirrors has shown a quality factor (Q-factor) reaching $10^6$ at cryogenic temperatures~\cite{Manenti2016}.  Moreover, a localized SAW mode in a high-quality SAW resonator as well as in a SAW waveguide can couple to different physical systems through the strain on the substrate.  Strong couplings with a superconducting qubit as well as a nitrogen vacancy center (NV center) are realized~\cite{V.Gustafsson2014,Golter2016}.  Couplings with other physical systems including quantum dots and trapped ions are also investigated theoretically~\cite{Schuetz2015}.  While there are many efforts for a practical use of the coupling of SAW with various physical systems in the radio-wave and microwave regimes, the investigation of the SAW coupling with optical systems remains scarce.  Stimulated optical excitation of SAWs on a microsphere resonator has been observed~\cite{Bahl2011}.  Applications of a propagating SAW as a coherent bus for the bulk acoustic resonator~\cite{Balram2015} and as a modulator for the nanophotonic optical cavity are reported~\cite{Li2015}.

In this article, we present an optomechanical system, where a planar SAW resonator is inserted perpendicularly to the light mode of an optical Fabry-P\'{e}rot cavity. Similar to the ``membrane in the middle''-type optomechanical system~\cite{Thompson2008}, the optoelastic effect induced by the strain in the SAW resonator effectively alters the cavity length~\cite{Matsko2009,Shumeiko2016}. The focusing effect in the two-dimensional (2D) SAW resonator also enhances the optomechanical coupling. This configuration additionally opens up a utility of the polarization degree of freedom in the optomechanical system, naturally introduced by the anisotropy of the SAW substrate and the tensorial nature of the optoelastic coupling.
 
\begin{figure*}[t]
 \centering
 \includegraphics[width=12.2cm]{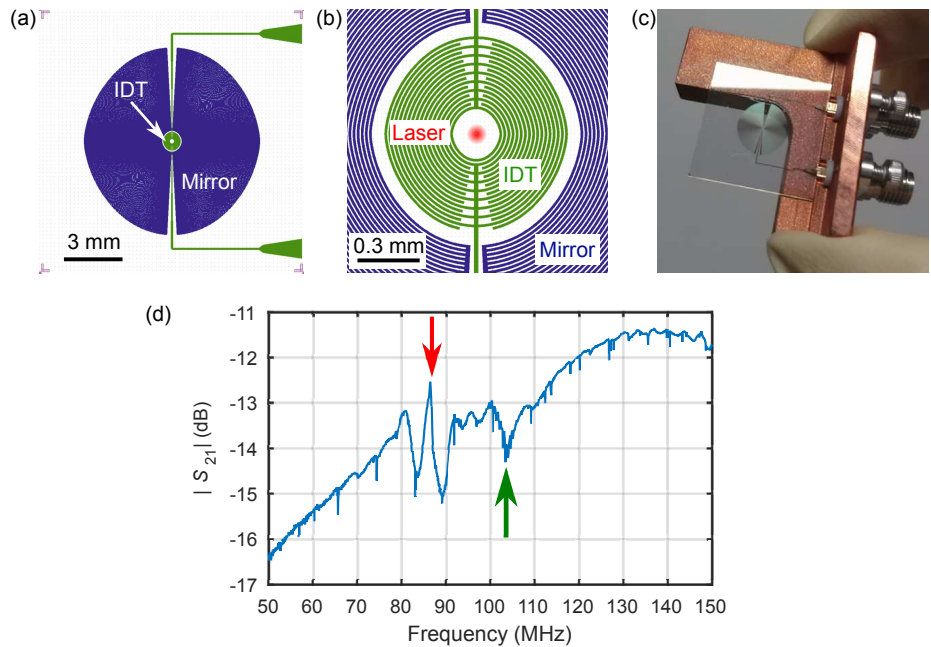}
 \caption{Two-dimensional SAW resonator. (a)~Design of the SAW focusing circuit. The outer~(blue) and inner~(green) patterns  form the concentric Bragg mirrors and interdigitated transducer~(IDT), respectively. (b)~Magnification of the center part. The width and spacing of the electrodes are 10~$\mu$m.
In order to map out the spatial variation of the optoelastic effect, a laser beam is scanned across the central region free from electrodes. (c)  Photograph of a SAW resonator. The chip is mounted on a sample holder and wirebonded to the SMA connectors.  Aluminum electrodes are evaporated on Y- and $128^\circ$Y-cut LiNbO$_3$ substrates. (d)~Radio-frequency transmission spectra through the SAW resonator.  A clear SAW resonance peak is observed at 87~MHz (red arrow) with a linewidth of 1.7~MHz.  A dip around 105~MHz is attributed to a leaky SAW mode~(green arrow).  Other spurious spikes are the overtones of the longitudinal and shear modes of the bulk waves.  All the materials shown in (a)-(d) are of the Y-cut sample.
}
 \label{figure1}
\end{figure*}%

\begin{figure}[t]
 \centering
 \includegraphics[width=8.52cm]{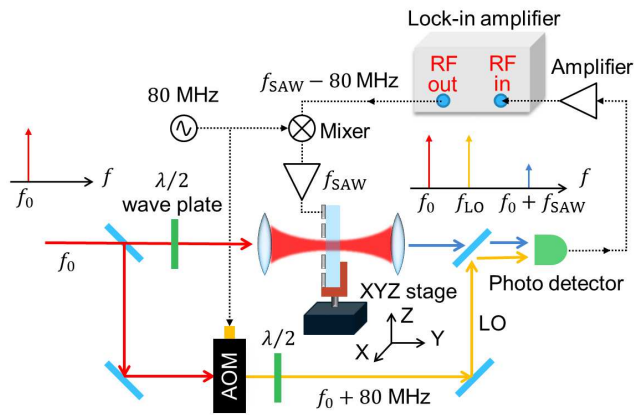}
 \caption{Experimental setup for probing the optical modulation imparted by the SAW induced optoelastic effect.  An RF signal of around 87~MHz is applied to the device for the SAW excitation.  A focused laser beam is sent through the SAW device.  A heterodyne signal with a local oscillator (LO) laser is detected and demodulated with a lock-in amplifier to observe the phase modulation.  The laser beam width at the SAW device is estimated to be 3.5~$\mu$m, much smaller than the SAW wavelength of 40~$\mu$m.  The SAW device is attached to an XYZ-stage, which allows observing the spatial distribution of the phase modulation.}
 \label{figure2}
\end{figure}%

\section{Experiment}
\subsection{Fabrication and RF spectroscopy of SAW resonators}
We fabricate 2D SAW resonators on $\rm{LiNbO_{3}}$ substrates with two different cuts, Y-cut and $128^\circ$Y-cut, whose thickness are 500~$\mu$m and 1000~$\mu$m, respectively.  The substrates are covered with anti-reflective (AR) coating on both sides.  The AR coating is a single layer $\rm{SiO_{2}}$ with thickness  of $183$~nm, much smaller than the SAW wavelength of tens of microns, chosen to mitigate the effect on the SAW properties.  The SAW resonator circuits are fabricated using photolithography and subsequent wet etching processes.  A 100-nm thick aluminum is evaporated on one side of the substrate and is patterned to form a concentric interdigitated transducer (IDT) and Bragg mirrors.  Due to the crystal anisotropy, the group velocity of the SAW strongly depends on the propagating direction.  We construct an anisotropic IDT, whose contour is proportional to the angular dependence of the group velocity for each cut.  Similarly, anisotropic Bragg mirrors are designed to focus the SAW at the center~\cite{Laude2006,Laude2008}.  
CAD images of the sample are shown in FIG.~1(a) and (b).  In both samples, the width and the spacing of the electrodes in the IDT and the Bragg mirrors are 10~$\mu$m, respectively, which excites the SAW with a wavelength of $\lambda_{\rm SAW} \simeq 40~\mu$m.  For this width, the Y-cut and $128^\circ$Y-cut SAW resonators are expected to have the resonance frequencies of $87.2$~MHz and 99.9~MHz, with corresponding SAW velocities of 3488~m/s and 3997 m/s, respectively~\cite{Campbell1998}.  The samples are fixed with varnish on a sample holder and the IDT ports are wire-bonded to SMA connectors as shown in FIG.~1(c).

We perform radio-frequency (RF) transmission measurement ($S_{21}$ measurement) on the samples using a vector network analyzer.  We apply an RF signal through one of the IDT ports and measure a transmitted signal to the other end of the electrodes composing the same IDT.  Some fraction of the injected RF signal is directly transmitted through capacitive and inductive coupling between the IDT electrodes, while some is  via conversion into the SAW resonator mode.  Therefore, the resonant frequency of the SAW resonator appears as the sharp peak on a broad background as shown in FIG.~1(d).  Due to the interference with the directly transmitted signal, the resonance is not observed as a simple Lorentzian peak.  
From the spectroscopy, we find the SAW resonance at 86.4~MHz and 98.3~MHz, with corresponding Q-factors of $Q \simeq 50$ and 450, for Y-cut and $128^\circ$Y-cut $\rm{LiNbO_3}$ samples, respectively.  
The relatively low Q-factors of these samples are due to the anisotropy of the SAW propagation and the small misalignment of the Bragg mirror structure with respect to the crystal axis in the fabrication process.

\subsection{Sideband scattering due to the optoelastic effect and SAW focusing}
Next we perform an optical measurement~(FIG.~2), where the phase shift of the probe beam induced by the optoelastic effect in the SAW resonator is detected through a heterodyne detection.  The probe laser light at the wavelength of  $\lambda_{\rm opt}=1064$~nm is tightly focused near the center of the SAW resonator.  The RF voltage excites the SAW resonator mode, and the optoelastic phase modulation imparted by the SAW produces red and blue sidebands.  We use a lock-in detection scheme to obtain the amplitude of the sideband signal.  The heterodyne signal between the generated sideband light and the local oscillator beam is detected on a fast photodetector and sent to a lock-in amplifier (Zurich Instruments HF2LI), which enables the evaluation of only one of the sidebands.

The sample is scanned along the $XZ$ plane perpendicular to the propagation direction of the laser light.  The result of the 2D mapping is shown in FIG.~3(a), where the polarization of the laser beam is parallel to the $Z$ direction.  
We note that the coordinates are aligned with the crystal axes of the LiNbO$_3$ substrate.  From the modulation pattern, we clearly observe the focusing of the SAW at the center of the resonator.  
The observed fringe period reflects the standing wave patterns expected for $\lambda_{\rm SAW}=40~\mu$m.

\begin{figure}[b]
 \centering
 \includegraphics[width=8.52cm]{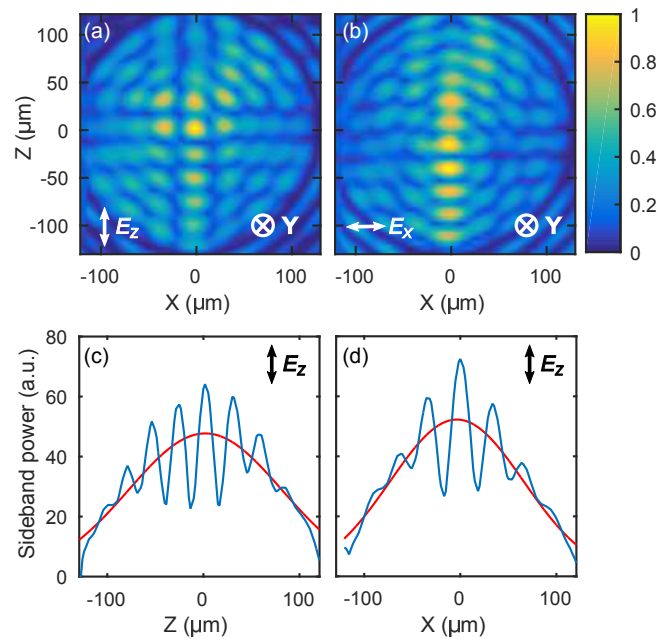}
 \caption{(upper panel)~Spatial distribution of the laser phase modulation imparted by SAW on the Y-cut LiNbO$_3$ sample.  Color maps show the normalized power of the generated sideband signal for the input laser of (a) vertical ($Z$-) and (b) horizontal ($X$-) polarizations, respectively.  The coordinates are aligned to the crystal axes of the sample.  (lower panel)~Cross sections (blue) attained by summing up the signal in FIG.~3(a) along the (c) $X$- or (d) $Z$- direction.  Gaussian fitting (red) to the data yields effective radii of (c) $R_Z = 110~\mu$m and (d) $R_X = 100~\mu$m as the $1/e$ decay length of the fitted Gaussian.
}
 \label{figure3}
\end{figure}%

\begin{figure*}[t]
 \centering
 \includegraphics[width=15.7cm]{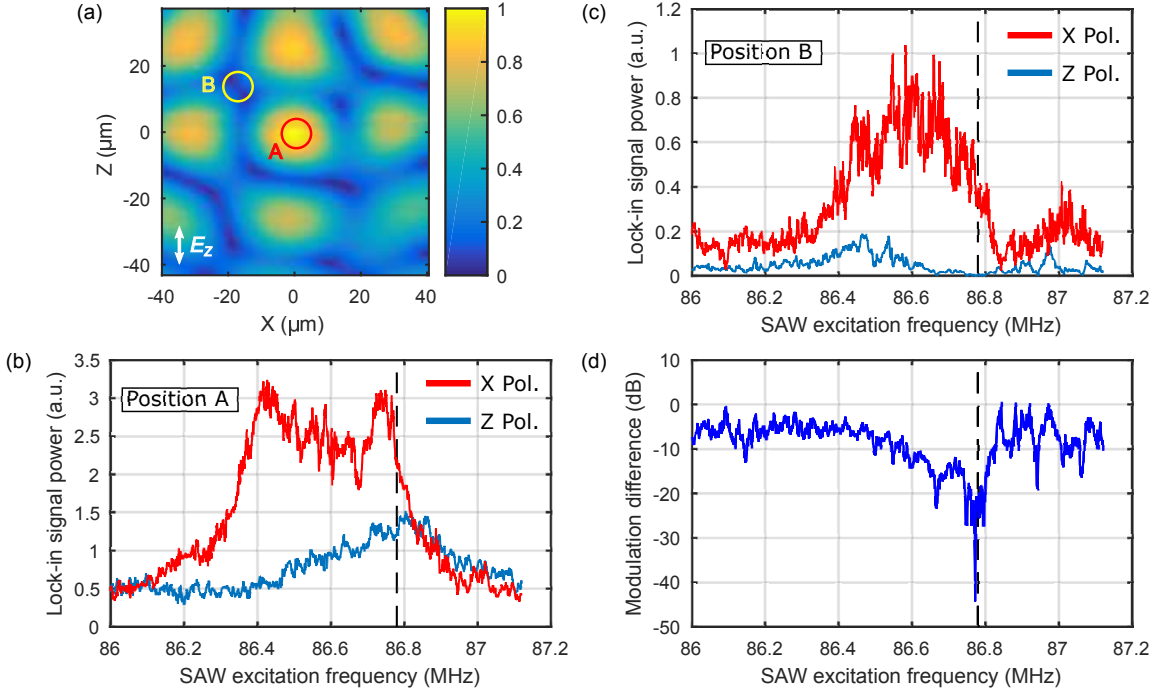}
 \caption{Polarization dependence of the SAW induced laser phase modulation.
(a)~Spatial distribution of the SAW resonator mode near its center, probed with the $Z$-polarized laser light.  Position~A, indicated by the red circle,  corresponds to the center node of the SAW resonator, where the maximum modulation is observed. Position~B~(yellow circle) is one of the antinodes.  (b)~Power spectra of the sideband generated by the $X$- (red) and the $Z$- (blue) polarized light at position~A.  (c) Same spectrum taken at position~B.  (d)  Ratio of the sideband power generated by the $X$- and $Z$-polarization light at position~B.  The black dashed lines in (b)-(d) indicate the SAW excitation frequency used in (a).  All the materials shown in (a)-(d) are of the Y-cut sample.
}
 \label{figure4}
\end{figure*}%

From the tensorial property of the optoelastic effect, one would expect different modulation patterns for different input beam polarization.  Changing the polarization of the probe beam to $X$ direction, with the SAW excitation frequency fixed, yields a quite different 2D image as shown in FIG.~3(b).  This observation indicates that the origin of sidebands generation is, in fact, due to the optoelastic effect and not the surface displacement.  
While the standing wave pattern is formed both in the $X$- and $Z$ directions in FIG.~3(a), that of FIG.~3(b) breaks the symmetry and is formed strongly in the $Z$ direction.

We can understand this observation qualitatively by considering the variation of the refractive index due to the optoelastic effect.  An optoelastic tensor contains eight independent coefficients $p_{11},\ p_{12},\ p_{13},\ p_{14},\ p_{31},\ p_{33},\ p_{41}$, and $p_{44}$ in the abbreviated notation \cite{Weis1985}.  Under the assumption that the $Y$ component of the displacement vector, $u_y$, is dominant in the optoelastic effect in the present setup, the relevant coefficients are only $p_{12}=0.088,\ p_{14}=-0.083$, and $p_{31}=0.177$ \cite{Jazbinsek2002}.  For the $Z$-polarized light it can be written as
\begin{equation}
\delta n_{z} = -\frac{1}{2} n_{e}^3 \, p_{31} \frac{\partial u_{y}}{\partial y}, \label{delnz}
\end{equation}
and for the $X$-polarized light as
\begin{equation}
\delta n_{x} = -\frac{1}{2} n_{o}^3 \left( p_{12} \frac{\partial u_{y}}{\partial y} + p_{14} \frac{\partial u_{y}}{\partial z} \right), \label{delnx}
\end{equation}
where $n_{e}=2.16$ and $n_{o}=2.24$ are the refractive indices for the extraordinary and the ordinary light, respectively.  Equation (\ref{delnz}) only contains the term ${\partial u_{y}}/{\partial y}$, an elongation rate along the $Y$ axis, whose distribution is visualized in FIG.~3(a), and under our assumption it may faithfully represent the SAW distribution in the resonator.  On the other hand, equation (\ref{delnx}) contains the additional term ${\partial u_{y}}/{\partial z}$, which causes the biased distribution toward the $Z$ axis and displaces the location of the maximum phase modulation along the $Z$ axis from $Z=0$ in FIG.~3(b).

A conventional SAW resonator with one-dimensional uniform standing wave has a simple mode area $A$, defined by the extent of the resonator.  The SAW focusing technique in the 2D concentric resonator allows one to reduce the mode area by concentrating the mode intensity at the focus of the resonator.  In such resonators, the spatial distribution of the SAW mode amplitude follows Bessel functions.  Integrating the circular area defined by the circuit, the mode has an effective area, $A=\pi R_{\rm eff}^2\,J_1^2(\alpha_{0 n})$, where $R_{\rm eff}$ is the effective radius of the circuit including a penetration depth into the Bragg mirror, $J_1$ is the first-order Bessel function, and $\alpha_{0 n}$ is the argument of the $n$-th zero of the zeroth-order Bessel function $J_0$, with $n \sim 2R_{\rm eff}/\lambda_{\rm SAW}$ being a number of nodes contained within the effective radius when $n$ is large.  For $R_{\rm eff}\gg \lambda_{\rm SAW}$ the expression simplifies to $A = \eta\,\lambda_{\rm SAW} R_{\rm eff}$, with $\eta \sim 0.32$.  The focusing in two-dimension allows the reduction of the mode area from the form $A = \pi R_{\rm eff}^2$ to $A = \eta\,\lambda_{\rm SAW} R_{\rm eff}$.  Using these expressions along with the effective radius of the original SAW resonator, $R_{\rm eff} \sim 1$~mm, and the wavelength $\lambda_{\rm SAW}=40~\mu$m, the mode area is expected to be reduced by a factor of 250.  In order to estimate an effective mode area of the 2D SAW resonator, we fit a Gaussian function to the SAW distribution as shown in FIGs~3(c) and (d).  The fit yields effective radii; $R_{Z} = 110~\mu$m and $R_{X} = 100~\mu$m, which is defined by $1/e$ decay length of the fitted Gaussian.  The effective mode area, expressed as $\pi R_{X}R_{Z}$, shows approximately 1/100 of the resonator's original mode area by means of the SAW focusing technique.

\subsection{Polarization dependent light-SAW coupling}
Utilizing the tensorial property derived above, we can selectively couple light to the SAW.  For example, for the beam spot focused at a node of the standing SAW mode in FIG.~3(a), the $Z$-polarized light is weakly modulated by the SAW, yet the $X$-polarized light is subject to strong modulation due to the second term of equation (\ref{delnx}).  To demonstrate this, we compare the depth of the phase modulation at two different positions, denoted A and B in FIG.~4(a), for different light polarizations.  First we set the beam spot at position A in FIG.~4(a) and sweep the RF frequency to obtain the sideband power spectrum for the $X$- and the $Z$-polarized light, shown as a red and a blue solid line in FIG.~4(b), respectively.  Next, we move to position B in FIG.~4(a), and perform the same measurement [FIG.~4(c)].  At position A each polarization is subject to the same amount of modulation imparted by the SAW mode. However, at position~B, $Z$-polarized light is hardly modulated while the $X$-polarized light still shows a sufficient modulation at the same frequency.  We note that the vertical axes of the two plots have the same scale.  FIG.~4(d) shows the ratio of the sideband power generated by the $X$- and $Z$-polarized light at position B.  At the SAW excitation frequency of 86.78~MHz, the modulation of the $Z$-polarized light is 300 times smaller than that of the $X$-polarization light, showing strong selectivity of light-SAW coupling depending on the polarizations.

\subsection{SAW-in-the-middle optomechanical system}
The observed optoelastic effect can be utilized to compose an optomechanical system.  We insert the 2D SAW resonator to an optical Fabry-P\'erot cavity, where the coupling between the SAW resonator and the optical cavity field via the optoelastic effect is investigated.  
The cavity is composed of two identical mirrors with the reflectivity of $\sim$99.5\%, and the radius of curvature of 25~mm.  The cavity length of about 50~mm, near the concentric configuration, is chosen to tightly focus the beam at the waist.  The unloaded cavity linewidth is measured to be 3.6~MHz, and we do not observe any measurable deterioration of the cavity finesse after loading the AR-coated SAW resonator chip inside the cavity.  The AR coating on the chip is quite important.  In our initial try without an AR coating, a large reflection of up to 15\% from each surface, owing to a large refractive index of $\rm{LiNbO_3}$, was enough to make the optical cavity unstable.

\begin{figure}[t]
 \centering
 \includegraphics[width=8.52cm]{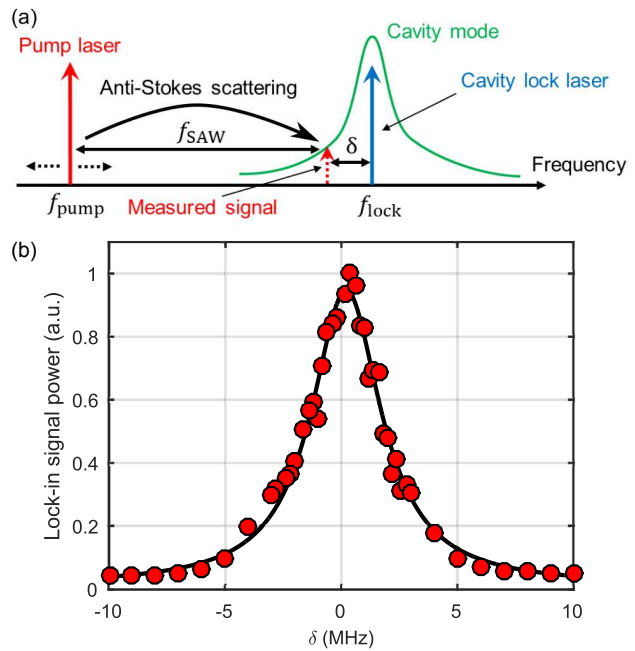}
 \caption{Cavity enhancement of the SAW-induced light scattering. (a)~Frequency diagram of the measurement. The pump laser frequency is set on the red side of the cavity resonance.  Anti-Stokes scattering of the pump laser is enhanced by the optical cavity mode when the detuning $\delta$ is close to zero.  The signal is obtained from the beat signal between the cavity lock laser and the scattered light by the SAW. (b)~Cavity enhanced spectrum.  The scattered light power shows a peak when the pump laser is swept across $\delta = 0$.  The observed linewidth of 3.6~MHz is consistent with the cavity linewidth measured independently. The result is obtained for the 128$^\circ$Y-cut LiNbO$_3$ sample.}
 \label{figure5}
\end{figure}%

With the SAW sample inside the optical cavity, we perform optical spectroscopy to observe the cavity enhanced anti-Stokes scattering by the SAW resonator.  The optical cavity is first locked with a reference laser via the Pound-Drever-Hall technique.  Next, while exciting the SAW by applying an RF signal to the IDT, we inject a strong pump power from the opposite side of the cavity under the detuning condition $\delta = f_{\rm lock} - \left(f_{\rm pump} + f_{\rm SAW}\right) \simeq 0$.  [See the frequency diagram in FIG.~5(a).]  The upper sideband of the pump laser scattered into the optical cavity mode is measured as a heterodyne signal between the up-converted signal and the lock laser reflected from the cavity.  
The cavity enhanced scattering is expected to be enhanced near $\delta \simeq 0$.  As shown in FIG.~5(b), we observe a Lorentzian-shape resonance, which shows significant enhancement of anti-Stokes scattering when the detuning from the cavity is tuned to the mechanical frequency of the SAW.  From the fit to the data the linewidth of the resonance is determined nearly identical to the cavity linewidth independently measured.  The cavity enhanced anti-Stokes scattering confirms that the SAW resonator is actually coupled to the optical cavity, composing a SAW-based cavity optomechanical system.

We analyze the system from the viewpoint of the cavity optomechanics to estimate the optomechanical coupling rate and other related parameters experimentally.  We focus the laser light at the center of the SAW resonator, while resonantly exciting the SAW by the RF with input power $P_{\rm SAW}$.  We first calibrate the phonon number stored in the SAW resonator, $N_\mathrm{SAW}$, from the RF reflection measurement ($S_{11}$ measurement).  Secondly, we measure the amount of the optical phase modulation, $\phi_{\rm SAW}$, imparted by the SAW by comparing the generated sideband power with that of a calibrated electro-optic modulator.  The phase modulation imparted by a zero-point fluctuation of the SAW in the resonator, $\phi_{\rm zpf}$, can be estimated as $\phi_{\rm zpf}=\phi_{\rm SAW}/\sqrt{N_{\rm SAW}}$.  The optomechanical coupling rate $g_0$ is defined as the shift of the cavity resonance frequency induced by a single phonon, $g_0 = (\omega_c/L)\,\delta x$, where $\omega_c$ is the resonance frequency of the optical cavity, $L$ is the cavity length, and $\delta x$ is the effective variance of the cavity length due to the zero-point motion of the SAW, in our case, written as $\delta x = (\phi_{\rm zpf}/2\pi)\,\lambda_{\rm opt}$.  For the Y-cut substrate, we find the optomechanical coupling rate of $g_0/2\pi = 60$~mHz.

A zero-point shear strain ${\partial u_{y}}/{\partial z} \simeq k_m U_{\rm zpf}$, where $k_m$ and $U_{\rm zpf}$ represent a wavenumber and an amplitude of the zero-point motion of the SAW, is experimentally obtained along with the equations (\ref{delnz}) and (\ref{delnx}).  Using the SAW wavelength of $\lambda_{\rm SAW}=40~\mu$m, we finally find $U_{\rm zpf} = 0.5 \times 10^{-2}$~fm, while a theoretical estimation derived from a value reported in \cite{Schuetz2015}, along with the effective mode area of the SAW resonator of $\pi \times 100~\mu {\rm m} \times 110~\mu {\rm m}$, yields $U_{\rm zpf} = 1 \times 10^{-2}$~fm.  The theoretical value is compatible with the experimentally estimated value, taking into account a large uncertainty accompanied by the calibration of $N_{\rm SAW}$.

\section{Prospects}
We have observed a clear indication of the SAW focusing resonator and the optomechanical coupling between the SAW and the optical cavity. However, the deterioration of the Q-factor of the SAW resonator, due to the velocity dispersion, and the small optomechanical coupling rate are serious problems to be tackled.  In order to obtain a high Q-factor in the focusing circuit, the SAW material constants, such as elastic and piezoelectric constants, from which we derive the group velocity distribution, need to be determined with high precision.  The inaccuracy of the electrode orientation with respect to the crystal axis in the fabrication process can be a deterioration factor as well.  In practice, it is not straightforward to achieve an extremely high Q-factor with LiNbO$_{3}$ and other materials with an anisotropy.  Instead, it is much easier to use an isotropic piezoelectric material such as GaAs or an non-piezoelectric isotropic substrate with an isotropic pioiezoelectric film such as AlN and ZnO, so that simple circular electrodes can be used to achieve the focused SAW whose Q-factor is limited by the internal loss and not from the velocity dispersion.

The small optomechanical coupling rate is mostly attributed to the large optical cavity mode volume and the small mode overlap between the optical and the SAW modes.  The optoelastic coupling is inversely proportional to the cavity length, $g_0 \propto 1/L$, where $L$ is the length of the optical cavity.  If a commercially available LiNbO$_3$ substrate with the thickness of 100~$\mu$m is used along with a fiber-optical cavity construction~\cite{Hunger2010,Flowers-Jacobs2012} with its cavity length of 300 $\mu$m, the optomechanical coupling rate of the order of 10~Hz could be achieved.  This coupling rate is comparable to that of the membrane-in-the-middle-type microwave-optical transducer whose conversion efficiency is nearly 10\% \cite{Andrews2014}.  While their transducer suffers from residual thermal phonons at 4~K because of the low mechanical frequency of the membrane, our transducer based on the high-frequency SAW, 100 times higher than in the membrane, could be cooled to the quantum ground state in the environment at 40~mK.  In addition, the rigid solid-state SAW device along with the fiber-based cavity would make the system more robust in the cryogenic environment.  The system's cooperativity defined by $C=4ng_0^2/\gamma \kappa$, where $n$ indicates intracavity photon numbers, $\kappa$ and $\gamma$ are decay rates of the optical cavity and the SAW resonator, respectively, would be increased up to $C \simeq 1$ after the modification described above, assuming the optical input power of 10~mW and the mechanical Q-factor of $10^5$, which we observe with a conventional planar SAW resonator on LiNbO$_{3}$ substrate.  In this regime, optomechanical control of the SAW resonator mode such as laser cooling would be possible.

For further improvement of the coupling strength, changing the coupling geometry would be desirable.  An optical waveguide embedded under a planar SAW resonator, where wave-vectors of both acoustic and optical waves are parallel, along with an appropriate phase matching condition, would ideally accomplish the perfect mode overlap between the acoustic and the optical field.

Our device opens up a new avenue in the application of the optomechanical system, where an optical polarization can be used as an additional degree of freedom.  In the context of quantum communication, the photon qubit basis which allows the heralded protocols, such as polarization or time-bin basis, are preferred.  Applications including polarization-to-Fock-state conversion between the optical and microwave signals may be possible.

\end{document}